# Making Models That Matter: How to Build Trustworthy and Useful Systems Biology Models

John M. Hancock[1]*, Mihail Anton[2,3], Frank T. Bergmann[4], Irina Balaur[5], Carissa Bleker[6], Thomas C. Collin[7], Oscar Dias[8,9], Elena Domínguez-Romero[10,11], Chris Evelo[10], Gavin Farrell[12], Martina Kutmon[13], Vitor Martins dos Santos[14,15], Anna Matuszyńska[16], Sébastien Moretti[17], Sara Morsy[18], Anna Niarakis[19,20], Marek Ostaszewski[21], Miguel Rocha[8,9], David Safranek[22], William Scott[23,24,25], Rahuman Sheriff[26,27,28,29], Silvio Tosatto[12], Dagmar Waltemath[30], Ulrike Wittig[31], Jan Zrimec[6], Anže Županič[6]

* Corresponding author. Email: john-michael.hancock@mf.uni-lj.si

[1] Faculty of Medicine, University of Ljubljana, Ljubljana, Slovenia
[2] Department of Life Sciences, Chalmers University of Technology, Gothenburg SE-412 96, Sweden
[3] ELIXIR, Wellcome Genome Campus, Hinxton, Cambridgeshire CB10 1SD, United Kingdom
[4] BioQUANT, Heidelberg University, Heidelberg, Germany
[5] Luxembourg National Data Service, Luxembourg
[6] Department of Biotechnology and Systems Biology, National Institute of Biology, Ljubljana, Slovenia
[7] Frontiers Media SA, Publishing Development, Lausanne, Switzerland
[8] Centre of Biological Engineering (CEB), University of Minho, Portugal
[9] LABBELS Associate Laboratory, Braga, Guimarães, Portugal
[10] Department of Translational Genomics (TGX), Maastricht University, Universiteitssingel 40, 6229 ER Maastricht, the Netherlands.
[11] Université de Lorraine, Laboratoire d'Ingénierie des Biomolécules (LIBio), Nancy F-54000, France.
[12] Department of Biomedical Sciences, University of Padova, Via Ugo Bassi 58/B, 35131 Padova, Italy
[13] Maastricht Centre for Systems Biology and Bioinformatics (MaCSBio), Maastricht University, The Netherlands
[14] Dept Bioprocess Engineering, Wageningen University and Research, The Netherlands,
[15] LifeGlimmer GmbH, Berlin, Germany
[16] Computational Life Science, Department of Biology, RWTH Aachen University, Germany
[17] SIB Swiss Institute of Bioinformatics, Vital-IT group, Lausanne, Switzerland
[18] Faculty of Health, Medicine, Life Sciences, University of Hertfordshire, U.K.
[19] Molecular, Cellular and Developmental Biology Unit (MCD), Centre for Integrative Biology (CBI), University of Toulouse, UPS, CNRS, Toulouse, France
[20] Lifeware Group, Inria, Saclay-île de France, Palaiseau, France
[21] Luxembourg Centre for Systems Biomedicine, University of Luxembourg

[22] Masaryk University, Faculty of Informatics, Brno, Czech Republic
[23] Department of Agricultural and Biological Engineering, Purdue University, West Lafayette, IN, United States
[24] UNLOCK, Wageningen University & Research, Wageningen, The Netherlands
[25] Laboratory of Systems and Synthetic Biology, Wageningen University & Research, Wageningen, The Netherlands
[26] European Bioinformatics Institute, European Molecular Biology Laboratory (EMBL-EBI), Cambridge UK
[27] Department of Surgery and Cancer, Faculty of Medicine, Imperial College London, London, UK
[28] Department of Medicine, Division of Pulmonary Systems Medicine, University of Florida, Gainesville, USA
[29] Earlham Institute, Norwich, UK
[30] Department of Medical Informatics, University Medicine Greifswald, Greifswald, Germany
[31] Heidelberg Institute for Theoretical Studies, Heidelberg, Germany

# Abstract

Computational models supporting mechanistic understanding of (complex) biological systems, systems behavior prediction, and experimental design are becoming more and more embedded in research on complex biological systems. Reuse and refinement of models, rather than continuous reinvention, is becoming increasingly important as models' demands on computational infrastructure increase. However published models - despite the variety of efforts taken so far - are frequently difficult to reproduce or reuse, substantially limiting their scientific value. Here we address the requirements for model reusability in the light of the field-specific CURE framework (Credible, Understandable, Reproducible, Extensible) and the more general FAIR principles (Findable, Accessible, Interoperable, Reusable). Considering published guidance we identify broad agreement on requirements for findability, accessibility, and interoperability, but continued lack of clarity and consensus around reusability. Focusing on the scientific quality and usability of computational models we discuss six key practices underpinning model sharing and re-use. Mapping the FAIR and CURE principles onto the model lifecycle we propose ten recommendations for building and sharing systems biology models that are both FAIR- and CURE-compliant.

# Introduction

Computational modelling of biological systems (systems biology) uses simplifications of biological systems to formalize our knowledge of them, and to predict their responses under varying conditions. This predictive capacity can guide experimental design, directly informing which experiments to prioritise and which parameters to measure next (Kitano 2002, Wolkenhauer *et al.* 2013, Voit 2022). As systems biology models become increasingly embedded into everyday biological research, a key issue becomes model re-use. In a recent white paper, the European ELIXIR Systems Biology Community identified two key barriers to the uptake of systems modelling in experimental contexts: the parameterization of published systems biology models is not always adequately reported, and there is a lack of standardisation and interoperability (Martins Dos Santos *et al.* 2024). Furthermore, model reproducibility, accuracy and reusability remain challenging. For example, 37% of 455 models in the BioModels database were found to be irreproducible in 2021, even after extensive curation efforts (Tiwari et al. 2021) and despite being encoded using COMBINE community standards (Hucka *et al.* 2015). Reproducible models in the same database were shown to have higher citation counts and impact than irreproducible models (Höpfl, Pleiss, and Radde 2022).

The issues surrounding model reproducibility, accuracy and reusability closely mirror the FAIR principles (Findable, Accessible, Interoperable, Reusable) originally developed for data (Wilkinson *et al.* 2016). A number of recommendation frameworks have been proposed to make models FAIR (Niarakis *et al.* 2021, 2022, Sauro 2021, Tiwari *et al.* 2021, Balaur *et al.* 2024, 2025). There have also been a number of Community-driven use cases developed to illustrate what is required to make models FAIR (Von Dassow *et al.* 2000, Cronin *et al.* 2023, Mendes 2023, Domínguez-Romero *et al.* 2024, Kruisselbrink *et al.* 2026).

The recommendations of these various deliberations and exercises are summarised in Table 1, in which similar recommendations from different publications are grouped together. We note that there is considerable consensus on what is required to make models more *Findable* and *Accessible*, and to a considerable extent *Interoperable*, but much less on what is needed to make them *Reusable*.

**Table 1**: Summary of published guidance on making systems biology models FAIR identifying main recommendations

| FAIR Principle | Application area in Systems Biology | Source |
|---|---|---|
| **Findable** | Data, model and documentation are stored on a public repository, shareable and with persistent identifiers.<br>Documentation includes model limitations.<br>Model versions are available via a version control system on a public repository | Sauro (2021)<br>Niarakis *et al*. (2022)<br>Cronin *et al*. (2023)<br>Dominguez-Romero *et al.* (2024)<br>Kherroubi Garcia *et al.* (2025)<br>Sauro *et al*. (2026)<br>Kruisselbrink *et al.* (2026) |
| | Metadata include the identifier of the model they describe and are registered or indexed in a searchable resource.<br>Metadata are structured and standardised, and their quality can be assessed. | Cronin *et al.* (2023)<br>Balaur *et al.* (2025a,b) |
| **Accessible** | Model code, description including biology, and data are available in a dedicated repository. | Sauro (2021)<br>Niarakis *et al.* (2022)<br>Sauro *et al.* (2026) |
| | Models are retrievable by their identifier using a standardised communications protocol, preferably free, and allowing authorisation. | Cronin *et al.* (2023)<br>Balaur *et al.* (2025a) |
| | Model (meta)data are accessible even when the model is no longer available. | Cronin *et al.* (2023)<br>Balaur *et al.* (2025a) |
| **Interoperable** | Models feature standard annotations for entities, interactions, and units.<br>Model annotations include qualified references to other objects. | Niarakis *et al.* (2022)<br>Cronin *et al.* (2023)<br>Balaur *et al.* (2025a)<br>Sauro *et al.* (2026) |
|  |  | Kruisselbrink *et al.* (2026) |
|  | Models and their (meta)data are described in a standardised manner.<br>(Meta)data use a formal, accessible, shared, and broadly applicable language for knowledge representation.<br>(Meta)data use vocabularies that follow FAIR principles. | Niarakis *et al.* (2022)<br>Cronin *et al.* (2023)<br>Dominguez-Romero *et al.* (2024)<br>Balaur *et al.* (2025a)<br>Kruisselbrink *et al.* (2026) |
|  | Models are interoperable with other software.<br>Model documentation indicates compatible software for model analysis. | Niarakis *et al.* (2022)<br>Cronin *et al.* (2023)<br>Dominguez-Romero *et al.* (2024)<br>Kruisselbrink *et al.* (2026) |
|  | Modelling software reads, writes and exchanges data in a way that meets domain-relevant community standards | Cronin *et al.* (2023)<br>Kruisselbrink *et al.* (2026) |
| **Reusable** | Models have a clearly defined scope, including the model objectives, assumptions, uncertainty quantifications, and applicability domain. | Loizou *et al.* (2008)<br>Dominguez-Romero *et al.* (2024)<br>Kherroubi Garcia *et al.* (2025)<br>Sauro *et al.* (2026) |
|  | Model and related (meta)data allow to repeat original results reported with the model. | Waltemath *et al.* (2011)<br>Sauro (2021)<br>Cronin *et al.* (2023)<br>Dominguez-Romero *et al.* (2024) |
|  | Documentation is provided on how to use the model code and data to generate the results reported with the model, including: i) description of equations or rules used in the model, ii) explicit mention of parameter values and units, iii) description of methodology. | Waltemath *et al.* (2011)<br>Sauro (2021)<br>Niarakis *et al.* (2022)<br>Cronin *et al.* (2023)<br>Dominguez-Romero *et al.* (2024)<br>Kruisselbrink *et al.* (2026) |
|  | Model code is provided with a series of formal tests, allowing to conduct code verification and model verification. | Sauro (2021)<br>Sauro *et al.* (2026)<br>Dominguez-Romero *et al.* (2024)<br>Kruisselbrink *et al.* (2026) |
|  | Models and their (meta)data support continuous integration combined with the formal tests to ensure that changes to the code do not produce errors. | Sauro (2021) |
|  | Modelling software and workflows follow software best practices, including implementing modular code.<br>Containerisation is used for simulation environments. | Sauro *et al.* (2026) |
|  | License of the model (re)use is clearly defined, allowing proper credits to the original version of the model. | Niarakis *et al.* (2022)<br>Cronin *et al.* (2023)<br>Dominguez-Romero *et al.* (2024)<br>Balaur *et al.* (2025a) |
| | Documentation follows domain-relevant community standards. | Cronin *et al.* (2023)<br>Balaur *et al.* (2025a)<br>Kherroubi Garcia *et al.* (2025)<br>Dominguez-Romero *et al.* (2024)<br>Kruisselbrink *et al.* (2026) |
| | Software includes qualified references to other software. | Cronin *et al.* (2023) |
| | (Meta)data are richly described with a plurality of accurate and relevant attributes | Andersen *et al.* (1995)<br>Cronin *et al.* (2023)<br>Dominguez-Romero *et al.* (2024)<br>Balaur *et al.* (2025a)<br>Kruisselbrink *et al.* (2026) |
| | (Meta)data are associated with detailed provenance, including model name and version, authors and their ORCIDs, source citation, and license. | Le Novere *et al.* (2005)<br>Dominguez-Romero *et al.* (2024)<br>Balaur *et al.* (2025a) |
| | Each relevant component of a simulation study has a unique ID. | Balaur *et al*. (2025a) |
| | Mathematical model description is consistent with the model code.<br>Mathematical models and their code are clear and understandable. | Le Novere *et al.* (2005)<br>Dominguez-Romero *et al.* (2024) |
| | Graphical conceptual model (used for visualization of the mathematical model) is consistent with the model code.<br>Model documentation and code use clear, concise, and consistent names for model components. | Andersen *et al.* (1995)<br>Dominguez-Romero *et al.* (2024) |
| | Modelling tools are user-friendly and support collaborative modeling practices. | Kherroubi Garcia *et al.* (2025)<br>Kruisselbrink *et al.* (2026) |
| | Model structure and described biological mechanisms are consistent, with documented choices of inferring rules or estimating parameters. | Niarakis *et al.* (2022) |
| | Involve the community in informing and promoting model-sharing practices.<br>Acknowledge diverse contributions.<br>Publicly recognize and reward research software engineers.<br>Influence publishers to promote good model-sharing practices.<br>Break down silos. | Kherroubi Garcia *et al.* (2025)<br>Kruisselbrink *et al.* (2026) |

Some authors have suggested that FAIR alone is insufficient for a broad view of best practice in model sharing. Domínguez-Romero *et al.* (2024) developed guidance to improve the understandability and reproducibility of physiologically-based pharmacokinetic (PBPK) models. Kherroubi Garcia *et al.* (2025) took a community-oriented approach to the sharing of models, emphasising transparency, collaboration, credit and recognition. A set of community guidelines designed to be complementary to the FAIR principles and specific to the systems biology domain has been framed under the acronym CURE (Credible, Understandable, Reproducible, and Extensible) (Sauro *et al.* 2026) to address the scientific quality and usability of computational models. Sauro *et al.* (2026) additionally proposed practical criteria and checklists to ensure that models are not only archived and accessible but also built on sound assumptions, transparently documented, validated against experimental evidence, and structured in a way that supports future extension and reuse in order to foster trust in computational biology, improve reproducibility, and support long-term model sustainability.

To synthesise this burgeoning diversity of guidance frameworks in systems biology, in this work we generalise the various FAIR recommendations and integrate them with CURE and other guidelines to provide concrete and streamlined recommendations to underpin the creation of trustworthy models that are applicable across the broad field of systems biology. For the purposes of this work, we use "trustworthy" to describe models for which the assumptions, design choices, data and evidence underlying their construction, implementation, and evaluation are sufficiently transparent and documented to allow users to understand, critically assess, reproduce, and appropriately reuse their results. We thereby aim to support the development and sharing of trustworthy and useful models.

In Figure 1 we present our conceptualization of the relationship of the four FAIR principles and the four CURE guidelines and how they are linked to four important concepts underlying trustworthy and useful models that are prominent in the literature summarized in Table 1: transparent design, detailed description, availability of code and simulation results, and the requirement for sensitivity analysis and model validation. Based on this conceptualization, in the following sections we present a set of considerations and best practices to address when developing, sharing (initial publication) and updating (subsequent improvements) system biology models.

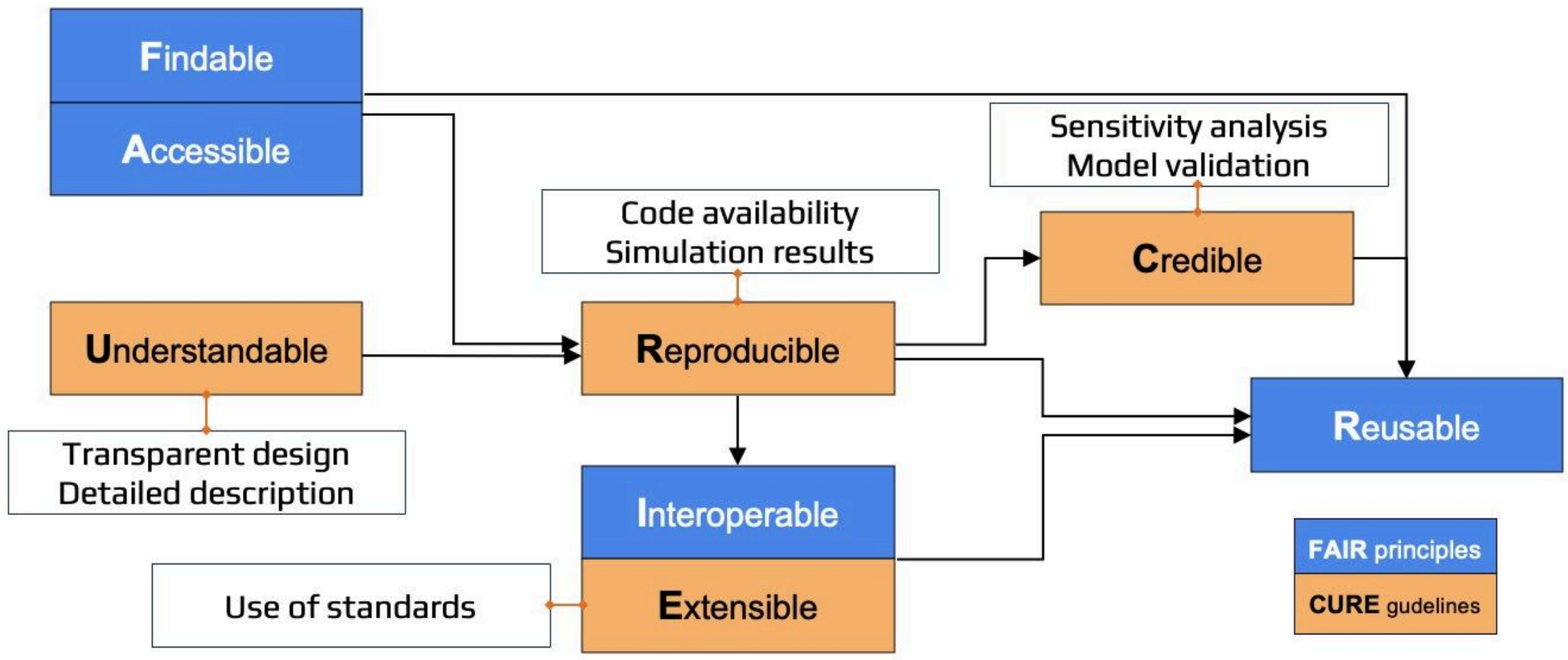

**Figure 1: Diagram of interrelations between FAIR principles and CURE guidelines in systems biology models.**

Representation of the interrelationships between FAIR principles (Wilkinson *et al.* 2016) , CURE guidelines (Sauro *et al.* 2026) and key concepts derived from Table 1 in this paper. FAIR and CURE concepts that are grouped are especially closely linked.

# Considerations for Trustworthy and Useful Models

In this section we discuss six considerations that need to be addressed for producing and sharing trustworthy and useful systems biology models. Subsequently we present ten recommendations and discussion points for preparing and sharing mechanistic computational models in biology. We take trustworthiness to be a consideration throughout the model lifecycle, rather than solely a property to be demonstrated when a model is prepared for sharing or publication. In particular, we emphasise that choices made during problem formulation and model design can determine whether a model can subsequently be understood, critically assessed, reproduced and appropriately reused.

# Developing an interpretable and understandable model

Having established that no published model meets their needs and developed a conceptual outline for the model to be developed, a systems biology modeller should ensure that the development process leads to an interpretable and understandable model. For this we recommend using modular and transparent design. This aligns with good software development practice (Artaza *et al.* 2016, Ousterhout 2021, Barker *et al.* 2022) and the CURE principles (Sauro et al. 2026), whereby credibility is inseparable from transparent, verifiable, and rigorously validated computational implementation.

## Modular development

Good software development practice can be extended to systems biology models by allowing for the separation of sub-models corresponding to individual sub-pathways, to encourage the re-use of components and reduce overheads of re-implementation.

With *modular design*, flexible systems are created by breaking them into independent, interchangeable parts (modules) with clear interfaces between them. This promotes easier development, maintenance, customization and reusability. The modular design principles include: (i) *abstraction* (each module internalises its complex internal workings, presenting its outputs to other modules via a simple, high-level interface), (ii) *encapsulation* (a module packages its components together, protecting its internal implementation to maintain its integrity), (iii) *cohesion* (each module performs a single, well-defined function, which contributes to its clarity and efficiency), (iv) *decoupling/independence* (modules operate separately with minimal dependencies on each other, allowing for easier changes and upgrades without affecting other parts of the system), and (v) *interchangeability* (modules can be independently created, modified, replaced, or exchanged with other modules within the same or different systems) (e.g. for knowledge graphs, Bleker *et al.* 2024; and for constraint-based metabolic models, Zrimec *et al.* 2025).

There are tools and standards that support the interoperability and reusability of such modular architecture. For instance, modular design is also supported by SBML level 3, allowing description of the modular architecture by providing a set of core features and an extensible package framework. These packages can help extend the core features by introducing or modifying different elements for model and submodel description, which allows reusability of the model (Keating *et al.* 2020). This version is also supported by different packages that enable the handling of different submodels like the Hierarchical Model Composition package (comp) (Smith *et al.* 2015). This can also help with transparent development, explained below.

## Transparent development

While modularity focuses on structural separation, transparency focuses on openness (visibility, clarity and predictability).

*Transparent design* emphasizes clarity and interpretability of internal workings, encouraging user understanding and interaction. Transparent design principles in the context of model development strongly overlap with modularity and include: (i) *visibility* (hierarchical model design encompassing internal structures and processes at different levels of complexity), (ii) *clarity and openness* (systems and processes are made understandable and accessible to users, avoiding hidden complexities or unnecessary barriers), (iii) *data provenance and lineage* (enabling tracking of the origin and path of data provides context and helps in understanding reliability and potential biases), and (v) *interpretability techniques* (formally stating or visualizing internal mechanisms can help users understand how decisions are made) (e.g., for knowledge graphs, Bleker *et al.* 2024).

Modular and transparent design work together, as transparency is facilitated by breaking a complex system into smaller, understandable and more easily modifiable components. These well-defined modules can contribute to overall system transparency, promoting clear documentation and leading to improved dissemination and communication, preventing implementation inconsistencies and ensuring long-term usability. Additional key practices derived from software best practice include: version control using git, readable code, and following a consistent coding style. An important aspect is documentation presented in the code itself (docstrings, comments) and in accompanying README files.

Use of approaches inspired by continuous integration pipelines can significantly help automatically run tests whenever the model or its accompanying code is modified. Since users can better understand the model's inner workings and how it functions, this also increases trust and improves the assessment of its reliability. When following these design principles, developers should consider the model's mathematical structure as well as implementation-specific data structures and algorithmic constraints. These are intrinsically linked to the type and level of possible modularity and transparency.

Externally, models can be thought of as possible modules in larger scale models that may be developed in future, so they should be implemented in such a way as to enable their reuse and capacity to plug in to other solutions (see the section on using standards to increase model interoperability). Moreover, due to constant software and package upgrades, it is sensible to think upfront about how to make a model usable over a longer period of time. Here, the safest route is either to use popular open-access solutions that are frequently updated and less likely to be outdated in the near future, or to use containerization to package model and software together, overcoming the problem of incompatibility with newer software versions. Another good practice that promotes clearer, more accurate and better implemented solutions, is to collaborate on the development (see for example Van Aalst *et al.* 2026). This way, the available expert knowledge in the community can be leveraged and expanded.

It should be noted that the use of AI tools in model development can hamper transparency because of black box components. For example, an AI application may have combined multiple sources to produce the model output: what are those sources and how did the AI algorithms combine them to give a specific output? Additionally, code generation has become increasingly accessible through the use of AI tools, but generating executable code does not establish that the implementation correctly represents the conceptual model or that the conceptual model is appropriate for its intended biological purpose. As minimal information, we recommend declaring as much information as possible about any AI tools used (e.g. AI tool and version used, last access date, modeling stages for which AI has been applied, further modifications made by the authors). We also recommend checking and citing the sources provided with the model output by the AI, in addition to the sources used by the authors. More broadly, the same FAIR and CURE principles applicable to human-built models should apply to AI-derived models.

# Verifying and validating the model

The requirement for model verification and validation aligns closely with robust software engineering practices (see previous section).

## Model verification

Model verification requires that the model code accurately implements the mathematical model and produces correct numerical results. In addition to high standards of software development this involves two complementary phases:

***Verification against suitable benchmark sets.*** This enables building confidence that the model behaves correctly. In particular, the model should be verified against related knowledge with known solutions to ensure that it provides the expected results. Only after a model passes these tests should it be trusted to produce credible results for new simulations that are not covered by the benchmarks.

***Assessing the usability of the model with concrete simulation or analysis methods.*** This is another important pre-condition of model credibility. This may include stiffness and time-step sensitivity analyses (for ODE, DDE, or PDE models), convergence tests for stochastic simulations, information on potential solver-induced variability (e.g., numerical methods stability or precision of the selected stochastic simulation algorithm), or checking update-scheme-dependent discrepancies in the case of qualitative models.

Some model deposition platforms, such as BioModels (Malik-Sheriff *et al.* 2019) and the Biodivine Boolean Models (BBM) repository (Pastva *et al.* 2023) directly support model development best practices. In the case of BioModels the support of syntactic- and semantic-level checking at the level of SBML operationalises software-engineering principles for model verification. BioModels also implements assessment of whether the MIRIAM (Le Novère *et al.* 2005) recommended, identifiers.org URL-based model annotations are resolvable. BBM performs systematic model consistency checks, including syntactic and semantic validation, and automated repair of common modelling inconsistencies. In the PBPK modeling field, the *FAIR PBK inspector* helps apply the new *FAIR PBK standard* and its harmonised annotations during model development and refinement. This helps increase model interoperability and reusability (Kruisselbrink *et al.* 2026). These model repositories and tools function as reliable benchmark suites for tools and algorithms operating on particular types of models. Their design illustrates how software development best practices directly support the CURE pillars of *credibility*, *reproducibility*, and *extensibility.*

### Model validation

Distinct from model verification, model validation assesses whether the model reproduces relevant biological observations or can be rigorously related to biological hypotheses. It involves comparing the outputs of the model with experimental outcomes in real biological systems (*in vitro* or *in vivo*). Model validation requires comparison with experimental data not used during development of the model (Andersen, Clewell, and Frederick 1995, ASME 2018, FDA 2021). In this context it is important to match simulation conditions with experimental conditions (initial states, growth medium, perturbations, etc.). An important point to note is that all model outputs may not be measurable in biological experiments for technical reasons. To address this it is important to consider when building models what the likely *in vitro*/*in vivo* measurements and other outputs describing experimental conditions may be, and to accommodate them in the model predictions. Finally, it needs to be remembered that validation is inherently context-dependent; a model validated for one type of prediction (e.g., steady-state fluxes) may not be valid for another (e.g., time-course dynamics).

Overall, credibility of models must be evaluated relative to a clearly specified context of use (applicability domain), targeting the following questions:

- What question is the model intended to answer?
- What phenomena does it explicitly represent, and what does it omit? What assumptions or simplifications are acceptable?
- Which parameter ranges and environmental conditions are within scope? For which conditions has the model been validated?

From a dissemination perspective, authors should seek out journals in which these considerations are taken seriously within reviewer guidelines and editorial resources, and where clear answers to such questions are recognized as integral to a rigorous assessment of the work. Model documentation should include explicit justifications for design choices and identify boundaries beyond which the model should not be applied.

## Using standards to increase model interoperability

Model interoperability allows models to be exchanged, reused, integrated, and executed across computational tools and platforms. Standards exist in systems biology to facilitate interpretation and seamless interoperability across different software platforms, preventing researchers from being trapped in tool-specific silos. Therefore, standards promote reproducibility, by unifying how models and related metadata are described. They also enable straightforward comparison of models from different origins, which in turn helps developers to avoid effort duplication when developing models, and to select the most suitable model for a specific task when several are available. Given the complexity and heterogeneity of biological systems, ensuring that models can interact seamlessly with diverse software environments and datasets is essential for advancing reproducible and collaborative research.

Different standards are useful in different contexts. SBML (Hucka *et al.* 2003) and CellML (Cuellar *et al.* 2003) are used to exchange models, including for submission to repositories such as BioModels. SBML, mainly used for representing biochemical reaction networks, has also been recommended as an exchange format for PBPK models (Kruisselbrink et al., 2026), while CellML is often used in multiscale physiological models. These formats are supported by numerous simulation and analysis tools, including COPASI (Hoops *et al.* 2006) and VCell (Blinov *et al.* 2017). An extensive list of simulation tools may be found in Cook *et al.* (2026). MIRIAM (Minimum Information Requested in the Annotation of Models) (Le Novere et al., 2005) is a set of guidelines that outline the essential metadata required for the annotation of computational models, improving understandability and traceability. SED-ML (Waltemath, Bergmann *et al.* 2011) implements the MIASE (Minimum Information About a Simulation Experiment) guidelines (Waltemath, Adams *et al.* 2011), thereby providing information about simulation experiments. COMBINE (Hucka *et al.* 2015, Waltemath *et al.* 2020), a community network developing standards and formats to share computational models in biology, provides a recommended collection of computational modeling standards (also to be found at the FAIRsharing resource https://fairsharing.org/FAIRsharing.ad3f2b) (the FAIRsharing Community *et al.* 2019).

To facilitate reproducibility checks by peers, it is necessary to provide precise simulation information and quantitative results for comparison, as per MIASE guidelines (Tiwari et al., 2021; Waltemath et al., 2011; Domínguez-Romero et al., 2024). BioModels provides a 9-point scorecard to help assess model reproducibility (Tiwari *et al.* (2021) (https://www.ebi.ac.uk/biomodels/reproducibility). SED-ML provides a generic route to reproducibility by encoding the procedures for running simulation experiments, including specification on which models, parameters, simulation procedures, and output plots are used. By either updating or newly adding SED-ML files systematically to models held in the BioModels database, Smith *et al.* (2025) could reproduce results of 88% of tested models using at least two different simulation engines. However different types of models entail different reproducibility challenges and may require specialised approaches. For genome-scale metabolic models (GEMs) and other constraint-based models Raman *et al.* (2024) introduced FROG (Flux variability, Reaction deletion, Objective function, Gene deletion) analysis. For Stochastic Simulation Studies, Sego *et al.* (2025) developed EFECT (Empirical Characteristic Function Equality Convergence Test). For logical models of gene regulatory networks (Boolean networks) automatic checking for logical consistency of models (i.e., normalisation, validation, and repair) is available via the Biodivine database (Pastva et al. 2023).

As general guidance we recommend applying the reproducibility scorecard (Tiwari *et al.* 2021), SED-ML encoding, and FROG, EFECT or Biodivine analysis if appropriate, to maximise model repeatability. Use of these scorecards and analyses should be clearly described during dissemination, so that reviewers can appropriately validate models and assess whether they are genuinely repeatable.

For effective dissemination, authors should clearly state which standards, file formats, and annotation frameworks have been used, so that the interoperability, reuse, and practical applicability of the model can be readily assessed by other researchers. Interoperability also extends to input and output data formats. Commonly used formats such as CSV or JSON allow structured data storage of raw data and data exchange between modelling tools and data repositories. Model inputs and outputs should comply with standard nomenclatures and their units should always be declared, and arbitrary numerical inputs or outputs should be avoided. As an example, Kruisselbrink *et al.* (2026) have proposed harmonised annotations for PBPK model components and units.

# Sharing Model Code

Model sharing is needed for model reuse and also addresses the FAIR principles of Findability and Accessibility: the reachability, retrievability, usability and visibility of models. Accessibility is essential for model re-use and validation (Sauro 2021, Niarakis *et al.* 2022, Cronin *et al.* 2023, Sauro *et al.* 2026). Once users identify a model, they need to know how it can be accessed.

Models should be stored in a dedicated, open-access model repository that assigns a persistent identifier (PID), such as a DOI, to the deposited model. Use of a dedicated repository is strongly recommended because sharing via standard websites, including journal websites, suffers from numerous disadvantages:

- Links can be broken so that the model is no longer available. 25-30% of links to systems models or bioinformatic tools are broken, with link rot worsening with time (Mangul *et al.* 2019, Ősz *et al.* 2019, Kern, Fehlmann, and Keller 2020).
- Links can point to a paywalled interface, or an interface with authentication and authorisation that users may not have. A search on the PubMed database for “Systems Models” indicates that of those published between January 2000 and April 2025 (n=1,428), only 28.7% of papers were published as fully open access. A further 25.4% were permanently paywalled, accessible only to readers with a journal subscription. The remaining 43.1% were published under a hybrid model, in which traditionally many authors pay for open access, but the article is typically embargoed behind a paywall for 12 months before becoming freely available, by which point it may have lost much of its novelty.

Submitting to a dedicated repository also makes the model automatically accessible, avoiding situations in which developers are no longer able or willing to respond to direct requests from the authors.

For dissemination purposes, repository deposition should be treated as part of the scholarly record itself, with the article, deposited model version, license, and PID linked bidirectionally. This ensures transparency and allows citation and long-term accessibility of the models. The TRUST Principles for digital repositories (Transparency, Responsibility, User focus, Sustainability and Technology) (Lin *et al.* 2020) provide a framework against which to evaluate trustworthy repositories. Model repositories typically contain a version of the model in a standard format and model metadata of greater or lesser detail. Model repositories differ in specialization and functionality and include specific metadata and domain-specific features. The choice of repository depends on the individual user and the model type. We provide a compilation of suitable model repositories via a FAIRSharing collection at https://fairsharing.org/6576.

The choice of the repository for sharing a model may also depend on the requirements of the journal in which it is published. Conversely these requirements may influence journal choice. Publication in online or open-access journals has some benefit as they are less constrained by traditional page limits and are therefore better positioned to support fuller reporting and dissemination of code and associated materials. Print-oriented publishers operating under stricter space limitations may be less likely to accommodate this approach.

Choice of repository will also be affected by considerations on the openness of the license to be adopted. For example, the BioModels database applies a CC0 license, which means that models can be freely downloaded and reused without obligation to attribute the original (https://creativecommons.org/publicdomain/zero/1.0/deed.en). An alternative, widely used open licence is Creative Commons Attribution (CC BY), which permits sharing and adaptation, including commercial reuse, provided attribution is given. This licence is preferred or required in major open-access policy frameworks such as Wellcome Trust and Plan S, and is consistent with the EU's broader emphasis on reusable scientific information. In some cases, Creative Commons Attribution-NonCommercial (CC BY-NC) may be preferred where restricting commercial reuse is important; however, because it limits reuse to non-commercial contexts, it may not align with some funder or dissemination-route requirements and should therefore be selected only after checking the policies of the intended journal, repository, and funder. In general submitters should select a licence that is "as open as possible" so that the model has minimal privacy and proprietary constraints, and is not under a licence inhibiting its accessibility. Authors should avoid a situation in which a model is in a non-usable or binary format, e.g. in a commercial format, so that users do not have the necessary licence to use it.

Options complementary to the use of model repositories are also available for sharing code. An elegant approach, especially for updating models initially submitted to dedicated repositories but which subsequently undergo updating, is to share code and documentation via an open distributed version control system such as git. Git repositories streamline the community curation process that often takes place after publication. Code repositories enable introducing changes, annotations and updates in a structured, explicit, and trackable manner, thereby facilitating community-driven efforts and cumulative science. Tags in git repositories are a good way to label specific versions of models so that users can always refer to the right one. Model deposition in git repositories can be facilitated by standardised templates adapted to the relevant platform that guide the modeler to ensure required data and metadata on model and model performance are presented, for example Standard-GEM for deposition of genome scale metabolic models (Anton *et al.* 2023). Leveraging the Zenodo integration with GitHub, standard-GEM encourages modelers to obtain DOIs for their model releases. However a potential drawback of this approach is lower findability. This is less serious in a widely-used resource such as GitHub but developers should bear in mind where users of their model are likely to conduct searches.

Sharing code in a raw form has the disadvantage that the user needs to be able to compile and run the code. This becomes increasingly problematic as code ages ("Bit rot"). Shared code should be accompanied by information on the programming language and compiler version used, to allow for compatibility checks. Any other dependencies should also be identified. A useful addition to code sharing in a raw form is to provide a packaged version of the code using one of a number of software containerization options, such as Apptainer or Docker, which can provide a packaged copy of code and dependencies which will run across operating systems. An alternative, systems biology-oriented option is the COMBINE Archive (Bergmann *et al.* 2014). FAIRDOMhub (https://fairdomhub.org/) (Wolstencroft *et al.* 2017) provides a convenient means to organise and annotate modelling experiments and publish them within the platform using the SED-ML format. SED-ML files can be run on the BIOSIMULATORS (https://biosimulators.org/) platform, which provides access to a range of simulation tools and provides an option to export Docker images. Furthermore, the RO-Crate initiative (https://www.researchobject.org/ro-crate/) provides a generic solution to the packaging of research objects – data, code and documentation – and can provide links to all the component parts of a modelling experiment (Soiland-Reyes *et al.* 2022). RO-crates can be published in resources such as Zenodo and receive a DOI.

# Describing the model

To ensure model understandability and reproducibility, and thereby to facilitate reusability, computational models should be accompanied by a clear and detailed description of their structure and implementation. A good model description describes various aspects of the model in as clear a manner as possible using rich and meaningful metadata.

The metadata description should follow community standards such as the MIRIAM guidelines and best practices outlined in recent reviews (see Table 1). Model description should cover the structure of the model itself, including the mathematical equations involved, as well as the individual entities that make up the model, how the components are linked and the model parameters and variables. The elements we recommend for model description are listed in Table 2.

**Table 2.** Recommended elements to include in model description documentation.

| **Aspect of the model described** | **Overview** |
|---|---|
| Authorship and contributions | A list of authors and contributors, ideally provided, e.g., in the popular CRediT (credit.niso.org) taxonomy format. Full credit should be given to individuals who made a significant contribution to the development of the model |
| Type of model being described and the operational context in which it is implemented | The Mathematical Modelling Ontology is available for this purpose |
| Model structure | In case of complex models, hierarchical, modular model development or multi-agent system, a description of modules, submodels and relationships |
| Coding framework | Information on the programming language and compiler version used, to allow for compatibility checks. Any other dependencies should also be identified. |
| Tools use | Especially where AI tools have been used, details including AI tool and version used, |
| | last access date, modeling stages for which AI has been applied, further modifications made by the authors |
| Model lineage (sources) | Information on whether the model is based on an existing model. Here, the publication of the original model, if available, should be cited, and a link to the model repository or service provider name, or website if applicable, provided. |
| Equations | A complete list of the underlying equations. These may be provided in the supplementary material in case of a publication but it is preferable also to include them in the documentation. |
| Model components | Identify the model components, which may be proteins, RNAs, genes or higher-level entities such as organelles, cells or tissues. A key issue is the unambiguous description of model components. Preferably, model components should be linked to a recognised, unambiguous identifier (see FAIRSharing Collection [link]). |
| Biological context and purpose | The biological context in which the model was developed and is expected to be useful, citing the underlying biological knowledge, with a thorough description of the problem being addressed (model purpose), the applicability domain, and model assumptions (OECD 2021). |
| Use of standards | Which standards, file formats, and annotation frameworks have been used to assure model interoperability |
| Graphical Presentation | Where appropriate include diagrams that illustrate the model structure (Andersen et al., 1995). Ensure that the figure gives an accurate representation of the model. |
| Accessibility | Provide at least one PID at which the model can be accessed |
| Reuse | Either provide documentation on how to reuse the model within the model documentation or provide a link to a file that includes these details |
| Licensing | Definition of model code and data sharing terms via a license. |

From a dissemination perspective, the description file can serve not only as documentation, but also as a citable landing page for the model as a research object, linking the article, repository record, provenance, and subsequent versions in an unambiguous manner. The possible formats for this description typically include: (i) A peer-reviewed publication or an open-access preprint), (ii) model repository documentation formats, such as GitHub Pages and readme files, and (iii) other forms of textual documentation. In the latter case a PID/DOI should be obtained.

The model description should not (solely) be provided as a scientific publication or supplementary material to a publication, as these are static and not ideal for providing all the material required. It is nonetheless important that the shared model accurately reflects the model described in the publication.

The description should be linked to the publicly available version of the model by quoting its PID - ideally, this link should be bidirectional and unambiguous. It is better to provide the documentation by way of a free-standing document via sustainable and well-indexed repositories providing DOIs such as GitHub, Zenodo, Figshare, or Dryad. FAIRDOMHub (Wolstencroft *et al.* 2017) has been developed explicitly to provide access to data, models, protocols (SOPs) and model descriptions. Using sites of this kind and linking to them from deposited models if necessary provides the best available solution to findability. Providing documentation via an institutional or custom website is less sustainable, as they are often degraded when a researcher leaves the hosting institution and are not recommended: Balaur *et al.* (2024) recommend that model metadata should remain accessible even if the model is not. Additional transparency and versioning can be provided by hosting the documentation (and model) in a distributed version control system which supports publication and releases to the above-mentioned repositories (such as GitHub; see above).

In addition to the use of standard formats for model description and annotation, semantic interoperability can be achieved by using controlled vocabularies and ontologies to describe model components. These annotations enable precise and consistent interpretation of biological entities and processes across different models. A wide variety of identifier sources are available. To facilitate the selection of suitable identifiers we have developed a FAIRSharing collection listing potential ID sources for different model components (http://fairsharing.org/8014/).

In gray-box modelling (Schweidtmann, Zhang, and Von Stosch 2024) a system integrates mathematical models with black box models, e.g. statistical or AI components, often implemented using multi-agent frameworks. In these cases the AI component should be documented using established AI metadata and reporting standards. Where appropriate, standards such as Model Cards (Mitchell *et al.* 2018) for AI model documentation and machine-readable metadata schemas (e.g., MLCommons Croissant - https://mlcommons.org/working-groups/data/croissant/ for datasets and associated metadata) can be adopted to improve interoperability, reproducibility, and reusability. In addition to documenting the AI component itself, the metadata should explicitly describe its integration with the mathematical model. Documenting this link enables researchers to understand how the hybrid system operates, independently evaluate or replace the AI component, and reuse either the AI or mathematical model while preserving reusability and reproducibility.

Although extending metadata beyond the model description may be considered outside the scope of traditional model documentation, researchers are encouraged to adopt Responsible AI metadata standards where appropriate. For example, Croissant-RAI (https://docs.mlcommons.org/croissant/docs/croissant-rai-spec.html) extends the MLCommons Croissant specification with metadata for responsible AI, describing dataset provenance, fairness, transparency, governance, risk, and accountability. Incorporating such metadata can facilitate responsible reuse of AI components within gray-box models and support compliance with emerging responsible AI and research reproducibility practices. This can be also approached from the perspective of the model reliability and credibility (C of CURE) to make sure that responsible AI standards are incorporated where possible.

## Describing how to use the model

As well as providing the model itself, we emphasise the importance of providing guidance on how to use the model and how to implement it (in a computational sense) (Sauro (2021)). The ideal framework for this is to provide example inputs and outputs both in a written form and in an executable form via a custom SED-ML file. A simulation protocol should be provided that provides example inputs and outputs for the simulation and conditions used to generate them, to allow users to compare outputs with biological data and test instantiations and implementations. It should also include simulation conditions, including software/programming environment, algorithm, changes in parameters/concentration/states and any data normalization described to allow the user to reproduce the results of the simulation in their own computational environment. The environmental conditions within which the model is supposed to operate, e.g. media conditions, should also be included if this is a consideration in the model. This documentation also provides the opportunity to make explicit what units the inputs/outputs and parameters are measured in and how they are defined, which should not be considered to be obvious. A model provider can help potential users to re-use a model by testing the ability of the model to run on different platforms. In any case, the model provider should state on which platforms the model has been tested

Users are likely to re-use a model under different conditions (e.g. with different input values) to those described in the original publication. These may not fall within a range for which the model was originally constructed so it is important to describe the range of parameters and biological conditions under which the model is expected to be valid. This can be addressed by sharing the parameter range within which the model was evaluated by the authors.

In terms of the model life cycle, we emphasise the importance of testing the model during its development against a defined set of inputs and the expected outputs. This exercise is not only important for model development, but it is also needed to help the user ensure that the model they have instantiated provides the outputs they should expect. Finally, models are often updated after a time, but previous versions are likely to continue to be used. It is therefore important to ensure that the model corresponding to a given set of inputs and outputs continues to be available after a new version is produced.

# Discussion

## Recommendations

The previous sections present a broad set of considerations that should be considered during model development, from initiation to the time when the model is developed and needs to be shared and accurately described. Based on these considerations, we propose a streamlined set of recommendations for model construction and sharing, applicable across the field of systems biology and linked to the relevant FAIR and CURE principles. These are listed in Table 3 and summarised in Figure 2. Figure 2 contextualises the recommendations into the different phases of the model development life cycle. As our recommendations are intended for the broad field of systems biology, specific requirements arising from the use of sensitive personal data, such as in personalised systems medicine models including digital twins and virtual patients, are outside the scope of this work and have been addressed elsewhere (e.g. by Mayer and Golebiewski (2024). The recommendations we present are built on the assumption that a new model needs to be built. It is also important to consider the possibility that an existing model can be re-used. Cook *et al* (2026) present a helpful workflow for model re-use and for the construction of models as well as useful lists of relevant tools.

**Table 3**: Recommendations on building trustworthy and Useful Systems Biology Models: Ten priorities for FAIR and CURE-compliant models.

| | Recommendation | When | FAIR principles | CURE principles | See chapter |
|---|---|---|---|---|---|
| 1 | Use modular and transparent design | Before or during development | Interoperable, Reusable | Understandable, Reproducible, Extensible | Designing an interpretable and understandable model |
| 2 | Use standardized, open formats | | Interoperable, Reusable | Extensible | Using standards to increase model interoperability |
| 3 | Validate and test the model | | Reusable | Credible, Reproducible | Verifying and validating the model |
| 4 | Describe the model thoroughly, annotate assumptions and limitations | | Reusable | Understandable, Reproducible, Credible | Describing the model |
| 5 | Engage in community curation | | Findable, Reusable | Credible, Reproducible | Designing an interpretable and understandable model |
| 6 | Version and document all changes, e.g. using an open access code repository | | Reusable | Understandable, Reproducible, Credible, | Depositing the model code in a persistent repository |
| 7 | Assign persistent identifiers by depositing in a persistent open access repository | After development | Findable, Reusable | Credible | Depositing the model code in a persistent repository |

| 8 | Ensure open access to code and data | | Accessible, Reusable | Understandable, Reproducible | Depositing the model code in a persistent repository |
|---|---|---|---|---|---|
| 9 | Provide complete metadata | | Findable, Reusable | Understandable, Reproducible | Describing the model |
| 10 | Share simulation protocols and parameters | | Accessible, Reusable | Reproducible, Understandable | Ensuring reproducible model simulation results |

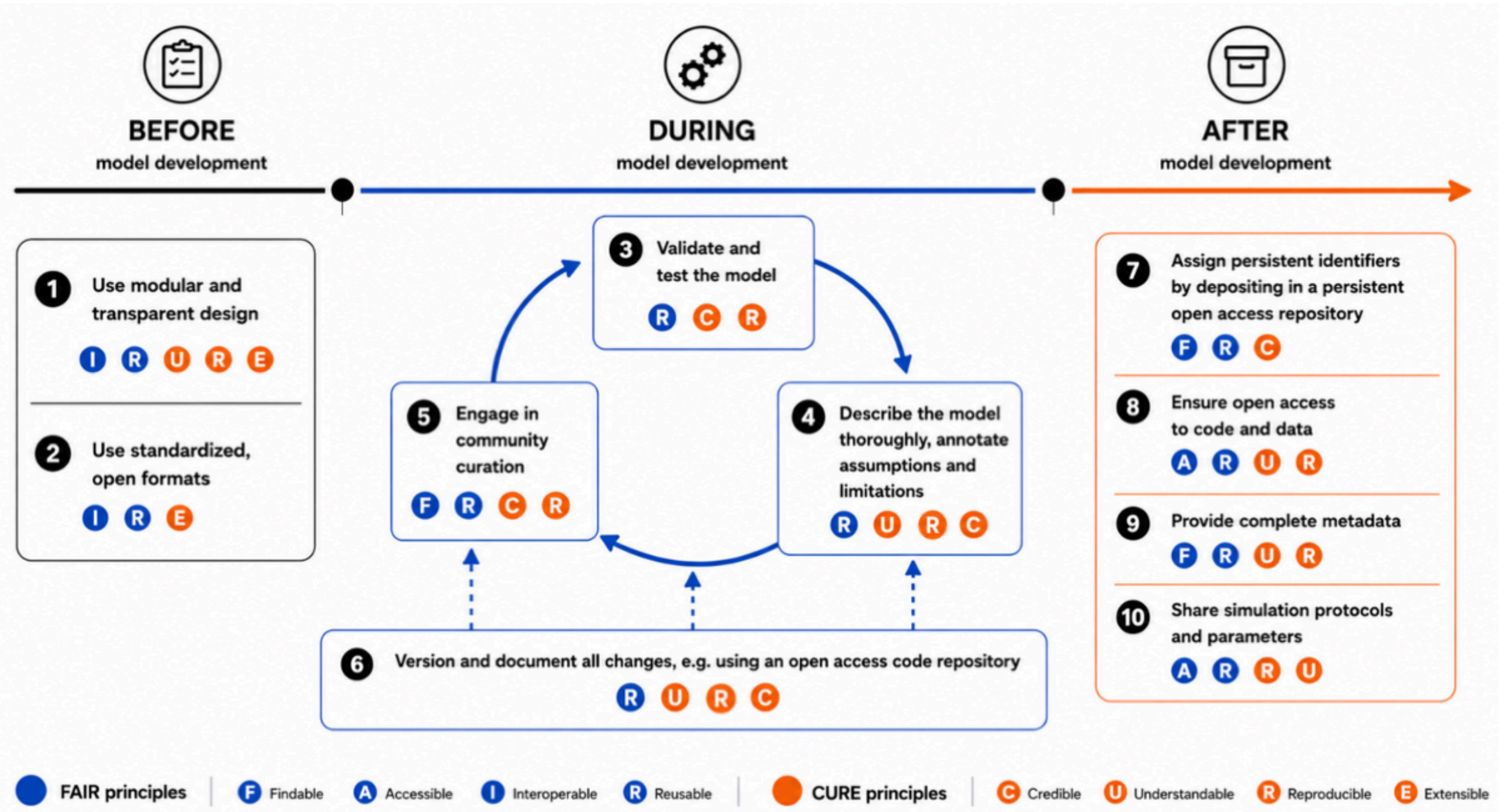

**Figure 2.** The ten recommendations mapped onto the model development lifecycle. Recommendations 1 and 2 should be addressed before during model development commences; recommendations 3-6 are relevant during model development, while recommendations 7-10 are relevant to model sharing after a shareable version has been finalised.

## Publishing and Disseminating Models

Open access in the context of FAIR and CURE should not be understood simply as free-to-read access to an article, but as persistent and reusable access to the model as a complete scholarly object consisting of the model in an open and standard format, the associated code, parameter sets, metadata, provenance, simulation protocols, version history, detailed description and clear licensing terms. This shifts a publication away from a static, article-centric record towards a linked research object model in which the article, model, code, data, documentation, and subsequent updates are explicitly connected and persistently citable. In a systems biology context, this shift is important because reproducibility, interpretability, and reuse depend not only on narrative description, but on whether the relevant computational artefacts can be located, executed, compared, and extended across platforms and over time.

For publishers, this has practical consequences for both editorial policy and publishing infrastructure. Requirements relating to model, code, and data availability may need to become more specific and more operational, with submission workflows that capture repository deposition, version-specific persistent identifiers, structured metadata, provenance, and clear reuse terms (licensing) at the point of submission. Likewise, peer review and editorial assessment may benefit from more explicit, FAIR/CURE-oriented guidance or checklists. This should not be seen as turning every reviewer into a technical curator, but rather as ensuring that minimum expectations for documentation, standards compliance, executable simulation workflows, and alignment between the manuscript and deposited artefacts are applied consistently. In this framework, publishers and editorial offices contribute not only to dissemination, but also to validation, curation, discoverability, and long-term stewardship, including the traceable citation of post-publication model updates and the maintenance of links between successive versions of the scholarly record.

Openness alone is insufficient. A model may be openly available and still remain effectively unusable if it lacks standard formats, rich metadata, clear provenance, executable workflows, or sustainable hosting. Equally, full openness will not always be appropriate where models depend upon sensitive, personal, proprietary, or commercially restricted material, in which case controlled access and clearly stated conditions of reuse may be more responsible than nominal openness. From a dissemination perspective, this also suggests that the contribution of publishers may increasingly lie less in access and more in providing the services that make reuse possible via curation, compliance assessment, interoperability support, trusted linking between articles and repositories, and long-term maintenance of the scholarly record. Licensing choices sit at the centre of this transition. More permissive licences may maximise downstream reuse, whereas more attribution-oriented approaches may better reflect scholarly credit expectations. In either case, the terms of reuse should be explicit, consistent with repository constraints, and legible to both human and machine users. A FAIR- and CURE-oriented publishing framework therefore depends not simply on openness, but on sustainable editorial and technical infrastructures that make models persistently accessible, intelligible, and reusable.

## Future Directions for the Systems Biology Community

An important area for future development in the systems biology community as a whole will be to address limitations of existing standards. As we have described, the community has developed a number of standards frameworks for transferring and describing models, but most of them have not been designed explicitly in light of the FAIR or CURE guidelines. (Exceptionally, the new *FAIR PBK* standard (Kruisselbrink et al. 2026) enables addressing interoperability and reusability of PBPK models.) More generally, there is a need to assess the different standards against FAIR and CURE and supplement or modify them accordingly. As these standards become more widely adopted, this will have the effect of improving FAIRness and CUREdness of models. Embedding CURE and FAIR within standards would also have the effect of raising recognition of these guidelines in the broader systems biology community. Whilst this paper is written from the perspective of a FAIR-aware community within systems biology, not all modellers are aware of these guidelines and how important they are for model sharing and re-use (and indeed they can sometimes be perceived as a burden).

Additionally, there is a need to evolve, adapt and enlarge current standards so that they can accommodate current and developing needs of the community. One example of such a need is multiscale and multi-component modelling (Sluka *et al.* 2016, Österberg *et al.* 2021, Schnitzer *et al.* 2022, Shameer *et al.* 2022), approaches that we believe can greatly benefit from a modular approach to modelling that would facilitate combining multiple models. Another example is gray-box modelling. Here, although MLDCAT-AP (Schiltz, Stani, and Weyts 2025) describes machine learning models, their datasets, quality measured on the datasets and citing papers, and OSAI defines how to derive sustainable, reusable and reproducible AI models (Farrell *et al.* 2026), these standards focus on ML or AI models alone.

Although systems biology throws up unique challenges that need to be addressed in its own standards framework, it will be important to learn from, and potentially align with, approaches to standards that are being developed in other fields. Examples are PRIMAD-LID (Aloqalaa, Soiland-Reyes, and Goble 2026) for computational reproducibility and standards for describing data origins in compliance with the Nagoya Protocol (Carroll *et al.* 2020, Raposo *et al.* 2026). The field must also recognise the need to take into account new developments, such as machine-learning-based approaches, by evolving and updating existing standards on a regular basis (Collins *et al.* 2024).

Artificial intelligence (AI) can be expected to play an increasingly important role in systems biology. At one level, AI tools can assist the development and publication of models in many ways, such as model validation and AI supported metadata filling, for example supporting standardised naming conventions. AI also has potential uses to address model understandability, for example to develop tutorials and guidance on model use, graphical representations of the relationships of entities within models, chatbots that can answer questions about models, and even spoken word interactions (Ruscone, Vazquez, and Valencia 2025, Wehling *et al.* 2025) although significant issues remain in using public AI tools to interpret models (Kannan *et al.* 2025).

Generative AI can be used to uncover hidden patterns and underlying principles from large, complex, and possibly unstructured data sources (Sengar *et al.* 2024), and produce realistic functional content, such as biological molecules and drugs (Repecka *et al.* 2021, Zrimec *et al.* 2022, Krishnan *et al.* 2025). We can therefore foresee that with increasing model complexity, data availability and AI development, AI will likely contribute more and more to conceptual aspects of modelling, revolutionising systems biology by shifting the focus from descriptive (knowledge-driven) modeling to (data-driven) generative design, thereby enabling *in silico* metabolic engineering (Docter, David, and Gohlke 2026), prediction of complex cellular behaviors (Zheng *et al.* 2025), and simulation of biological systems at multiple scales via the creation of virtual cells (Jiang *et al.* 2026, Thornburg *et al.* 2026).

Despite this exciting progress, models, whether created by humans or (partially) generated by AI, need to be understandable to humans and machine readable. Principles of transparency, accountability, and responsible use are fundamentally tied to ethics in computational modeling, especially as models increasingly inform high-stakes decisions in science, medicine, and policy. Predictive models, by their nature, simplify complex realities, and without proper oversight, they can perpetuate bias, obscure assumptions, and mislead stakeholders (Van Der Aalst, Bichler, and Heinzl 2017). Many of these concerns apply to AI-generated as well as manually generated systems biology models. Together, FAIR and CURE provide a foundation for ethical modeling practice by making models not only transparent and accessible, but also trustworthy and socially responsible. Black box models developed by AI, which are not human-understandable, run the risk of embedding potential bias and errors in model assumptions, structure or other aspects, that would be hard to detect and correct, biasing potential interpretations and conclusions obtained from those models.

The OSAI recommendations are a group of recommendations around reproducibility, environmental sustainability and reusability of AI models (Farrell *et al.* 2026). These recommendations are centred on clear documentation, FAIR repositories and tools and green development. The OSAI recommends that researchers report the environmental impact of their model as part of their model documentation. It also stresses the use of standardised benchmarking protocols for model evaluation, concepts that largely align with our recommendations. A key remaining question is how these concepts can be connected and aligned to the mathematical component of a systems biology model. For gray-box modelling that uses AI and mathematical modelling, descriptors describing how mathematical models and AI components operate together are therefore required, capturing the modularity of these models. More broadly we believe there is a need to facilitate alignment between community standards so that advances in one community can be adopted by another.

The framework we describe in this paper emphasises the importance of systems biology models as community resources and not solely as the outputs of individual research projects. Duplication of effort is a waste of resources and time, and FAIR and CURE-compliant model sharing serves to reduce duplication and helps the field advance more rapidly. Opening up models to input and curation by a wider group of contributors beyond the original core group of developers can further support this goal. We suggest that existing repositories develop functionalities to support the further evolution of existing models, rather than just reflecting the frozen state of submitted versions. In all cases, it is important to respect the licence terms under which the original models were published.

Finally, while this paper aims to summarise and integrate FAIR and CURE guidelines, we welcome broader community input to refine the guidance we have developed.

# Acknowledgements

This work was supported in part by the ELIXIR Implementation Study SYBEL: Systems Biology for ELIXIR.

# References

Aloqalaa M, Soiland-Reyes S, Goble C. PRIMAD-LID: A Developed Framework for Computational Reproducibility, version 1. Preprint, arXiv, 2026. https://doi.org/10.48550/ARXIV.2601.02349.

Alper P, D'Anna F, Droesbeke B *et al.* RDMkit: A research data management toolkit for life sciences. *Patterns* (New York, N.Y.) 2025;**6**(9):101345. https://doi.org/10.1016/j.patter.2025.101345.

Andersen ME, Clewell HJ, Frederick CB. Applying Simulation Modeling to Problems in Toxicology and Risk Assessment: A Short Perspective. *Toxicol Appl Pharmacol* 1995;**133**(2):181–7. https://doi.org/10.1006/taap.1995.1140.

Anton M, Almaas E, Benfeitas R *et al.* standard-GEM: standardization of open-source genome-scale metabolic models. Preprint, Systems Biology, 23 Mar. 2023. https://doi.org/10.1101/2023.03.21.512712.

Artaza H, Chue Hong N, Corpas M *et al.* Top 10 metrics for life science software good practices. *F1000Research* 2016;**5**:2000. https://doi.org/10.12688/f1000research.9206.1.

ASME. *Assessing Credibility of Computational Modeling through Verification and Validation: Application to Medical Devices: ASME V&V 40 - 2018*. New York: The American Society of Mechanical Engineers, 2018.

Balaur I, Welter D, Rougny A *et al.* MINERVA FAIR assessment fosters open science & scientific crowd-sourcing in systems biomedicine. Preprint, 29 Aug. 2024. https://doi.org/10.1101/2024.08.28.610042.

Balaur I, Welter D, Rougny A *et al.* FAIR assessment of Disease Maps fosters open science and scientific crowdsourcing in systems biomedicine. *Sci Data* 2025;**12**(1):851. https://doi.org/10.1038/s41597-025-05147-w.

Barker M, Chue Hong NP, Katz DS *et al.* Introducing the FAIR Principles for research software. *Sci Data* 2022;**9**(1):622. https://doi.org/10.1038/s41597-022-01710-x.

Bergmann FT, Adams R, Moodie S *et al.* COMBINE archive and OMEX format: one file to share all information to reproduce a modeling project. *BMC Bioinformatics* 2014;**15**(1):369. https://doi.org/10.1186/s12859-014-0369-z.

Bleker C, Ramšak Ž, Bittner A *et al.* Stress Knowledge Map: A knowledge graph resource for systems biology analysis of plant stress responses. *Plant Commun* 2024;**5**(6):100920. https://doi.org/10.1016/j.xplc.2024.100920.

Blinov ML, Schaff JC, Vasilescu D *et al.* Compartmental and Spatial Rule-Based Modeling with Virtual Cell. *Biophys J* 2017;**113**(7):1365–72. https://doi.org/10.1016/j.bpj.2017.08.022.

Carroll SR, Garba I, Figueroa-Rodríguez OL *et al.* The CARE Principles for Indigenous Data Governance. *Data Sci J* 2020;**19**:43. https://doi.org/10.5334/dsj-2020-043.

Collins GS, Moons KGM, Dhiman P *et al.* TRIPOD+AI statement: updated guidance for reporting clinical prediction models that use regression or machine learning methods. *BMJ* 2024;**385**:e078378. https://doi.org/10.1136/bmj-2023-078378.

Cook M, Anastasakis S, Heydarabadipour A *et al.* A standardized workflow for kinetic metabolic model curation and dissemination. *PLOS Comput Biol* 2026;**22**(4):e1014227. https://doi.org/10.1371/journal.pcbi.1014227.

Cronin MTD, Belfield SJ, Briggs KA *et al.* Making in silico predictive models for toxicology

FAIR. *Regul Toxicol Pharmacol* 2023;**140**:105385. https://doi.org/10.1016/j.yrtph.2023.105385.

Cuellar AA, Lloyd CM, Nielsen PF *et al.* An Overview of CellML 1.1, a Biological Model Description Language. *SIMULATION* 2003;**79**(12):740–7. https://doi.org/10.1177/0037549703040939.

Docter S, David B, Gohlke H. Deep learning and generative artificial intelligence methods in enzyme and cell engineering. *Curr Opin Biotechnol* 2026;**97**:103393. https://doi.org/10.1016/j.copbio.2025.103393.

Domínguez-Romero E, Mazurenko S, Scheringer M *et al.* Making PBPK models more reproducible in practice. *Brief Bioinform* 2024;**25**(6):bbae569. https://doi.org/10.1093/bib/bbae569.

Farrell G, Adamidi E, Andrade Buono R *et al.* Open and sustainable AI: challenges, opportunities and the road ahead in the life sciences. *Nat Methods* published online 20 Mar. 2026. https://doi.org/10.1038/s41592-026-03037-6.

FDA. Assessing the Credibility of Computational Modeling and Simulation in Medical Device Submissions. Food and Drug Administration, 2021. https://www.regulations.gov/document/FDA-2021-D-0980-0002 (30 Jan. 2026, date last accessed).

Hoops S, Sahle S, Gauges R *et al.* COPASI—a COmplex PAthway SImulator. *Bioinformatics* 2006;**22**(24):3067–74. https://doi.org/10.1093/bioinformatics/btl485.

Höpfl S, Pleiss J, Radde N. Bayesian hypothesis testing reveals that reproducible models in Systems Biology get more citations. Preprint, 7 Oct. 2022. https://doi.org/10.21203/rs.3.rs-2132474/v1.

Hucka M, Finney A, Sauro HM *et al.* The systems biology markup language (SBML): a medium for representation and exchange of biochemical network models. *Bioinformatics* 2003;**19**(4):524–31. https://doi.org/10.1093/bioinformatics/btg015.

Hucka M, Nickerson DP, Bader GD *et al.* Promoting Coordinated Development of Community-Based Information Standards for Modeling in Biology: The COMBINE Initiative. *Front Bioeng Biotechnol* 2015;**3**(19):doi: 10.3389/fbioe.2015.00019. https://doi.org/10.3389/fbioe.2015.00019.

Jiang H, Huang X, Bi X *et al.* Artificial intelligence-enabled multi-scale virtual cell: perspective, challenges, and opportunities. *Brief Bioinform* 2026;**27**(2):bbag104. https://doi.org/10.1093/bib/bbag104.

Kannan M, Bridgewater G, Zhang M *et al.* Leveraging public AI tools to explore systems biology resources in mathematical modeling. *Npj Syst Biol Appl* 2025;**11**(1):15. https://doi.org/10.1038/s41540-025-00496-z.

Keating SM, Waltemath D, König M *et al.* SBML Level 3: an extensible format for the exchange and reuse of biological models. *Mol Syst Biol* 2020;**16**(8):MSB199110. https://doi.org/10.15252/msb.20199110.

Kern F, Fehlmann T, Keller A. On the lifetime of bioinformatics web services. *Nucleic Acids Res* 2020;**48**(22):12523–33. https://doi.org/10.1093/nar/gkaa1125.

Kherroubi Garcia I, Erdmann C, Gesing S *et al.* Ten simple rules for good model-sharing practices. *PLOS Comput Biol* 2025;**21**(1):e1012702. https://doi.org/10.1371/journal.pcbi.1012702.

Kitano H. Computational systems biology. *Nature* 2002;**420**(6912):206–10. https://doi.org/10.1038/nature01254.

Krishnan A, Anahtar MN, Valeri JA *et al.* A generative deep learning approach to de novo antibiotic design. *Cell* 2025;**188**(21):5962-5979.e22. https://doi.org/10.1016/j.cell.2025.07.033.

Kruisselbrink JW, Minnema J, Deepika D *et al.* An exchange standard for FAIR PBK models in chemical risk assessment – A PARC community effort. *Comput Toxicol* 2026;**39**:100426. https://doi.org/10.1016/j.comtox.2026.100426.

Le Novère N, Finney A, Hucka M *et al.* Minimum information requested in the annotation of biochemical models (MIRIAM). *Nat Biotechnol* 2005;**23**(12):1509–15. https://doi.org/10.1038/nbt1156.

Lin D, Crabtree J, Dillo I *et al.* The TRUST Principles for digital repositories. *Sci Data* 2020;**7**(1):144. https://doi.org/10.1038/s41597-020-0486-7.

Loizou G, Spendiff M, Barton HA *et al.* Development of good modelling practice for physiologically based pharmacokinetic models for use in risk assessment: The first steps. *Regul Toxicol Pharmacol* 2008;**50**(3):400–11. https://doi.org/10.1016/j.yrtph.2008.01.011.

Malik-Sheriff RS, Glont M, Nguyen TVN *et al.* BioModels—15 years of sharing computational models in life science. *Nucleic Acids Res* 8 Nov. 2019:gkz1055. https://doi.org/10.1093/nar/gkz1055.

Mangul S, Martin LS, Eskin E *et al.* Improving the usability and archival stability of bioinformatics software. *Genome Biol* 2019;**20**(1):47. https://doi.org/10.1186/s13059-019-1649-8.

Martins Dos Santos V, Anton M, Szomolay B *et al.* Systems Biology in ELIXIR: modelling in the spotlight. *F1000Research* 2024;**11**:1265. https://doi.org/10.12688/f1000research.126734.2.

Mayer G, Golebiewski M. *Standardization Landscape, Needs and Gaps for the Virtual Human Twin (VHT)*, with Golebiewski M, Mayer G. Zenodo: Zenodo, 2024. https://doi.org/10.5281/ZENODO.10492795.

Mendes P. *Reproducibility and FAIR Principles: The Case of a Segment Polarity Network Model*, version 1. published online 2023. https://doi.org/10.48550/ARXIV.2304.08688.

Mitchell M, Wu S, Zaldivar A *et al. Model Cards for Model Reporting*, version 2. published online 2018. https://doi.org/10.48550/ARXIV.1810.03993.

Niarakis A, Kuiper M, Ostaszewski M *et al.* Setting the basis of best practices and standards for curation and annotation of logical models in biology—highlights of the [BC]2 2019 CoLoMoTo/SysMod Workshop. *Brief Bioinform* 2021;**22**(2):1848–59. https://doi.org/10.1093/bib/bbaa046.

Niarakis A, Waltemath D, Glazier J *et al.* Addressing *barriers in comprehensiveness,*

*accessibility, reusability, interoperability and reproducibility of computational models in systems biology*. *Brief Bioinform* 2022;**23**(4):bbac212. https://doi.org/10.1093/bib/bbac212.

Österberg L, Domenzain I, Münch J *et al.* A novel yeast hybrid modeling framework integrating Boolean and enzyme-constrained networks enables exploration of the interplay between signaling and metabolism. *PLOS Comput Biol* 2021;**17**(4):e1008891. https://doi.org/10.1371/journal.pcbi.1008891.

Ősz Á, Pongor LS, Szirmai D *et al.* A snapshot of 3649 Web-based services published between 1994 and 2017 shows a decrease in availability after 2 years. *Brief Bioinform* 2019;**20**(3):1004–10. https://doi.org/10.1093/bib/bbx159.

Ousterhout JK. *A Philosophy of Software Design*. Second edition, Palo Alto, CA: Yaknyam Press, 2021.

Pastva S, Šafránek D, Beneš N *et al.* Repository of logically consistent real-world Boolean network models. Preprint, Systems Biology, 12 June 2023. https://doi.org/10.1101/2023.06.12.544361.

Plesser HE. Reproducibility vs. Replicability: A Brief History of a Confused Terminology. *Front Neuroinformatics* 2018;**11**:76. https://doi.org/10.3389/fninf.2017.00076.

Porubsky VL, Goldberg AP, Rampadarath AK *et al.* Best Practices for Making Reproducible Biochemical Models. *Cell Syst* 2020;**11**(2):109–20. https://doi.org/10.1016/j.cels.2020.06.012.

Raman K, Kratochvíl M, Olivier BG *et al.* FROG Analysis Ensures the Reproducibility of Genome Scale Metabolic Models. Preprint, Systems Biology, 26 Sept. 2024. https://doi.org/10.1101/2024.09.24.614797.

Raposo DS, Faggionato D, Ebert B *et al.* How can biological databases support the new UN mechanism for benefit-sharing from digital sequence information? *Sci Data* 2026;**13**(1):971. https://doi.org/10.1038/s41597-026-07725-y.

Repecka D, Jauniskis V, Karpus L *et al.* Expanding functional protein sequence spaces using generative adversarial networks. *Nat Mach Intell* 2021;**3**(4):324–33. https://doi.org/10.1038/s42256-021-00310-5.

Ruscone M, Vazquez M, Valencia A. Intelligent Tool Orchestration for Rapid Mechanistic Model Prototyping: MCP Servers as AI-Biology Interfaces. Preprint, Bioinformatics, 15 Sept. 2025. https://doi.org/10.1101/2025.09.10.675105.

Sauro HM. The Practice of Ensuring Repeatable and Reproducible Computational Models, arXiv:2107.05386. Preprint, arXiv, 7 July 2021. https://doi.org/10.48550/arXiv.2107.05386.

Sauro HM, Agmon E, Blinov ML *et al.* From FAIR to CURE: guidelines for computational models of biological systems. *Npj Syst Biol Appl* published online 27 Mar. 2026. https://doi.org/10.1038/s41540-026-00651-0.

Schiltz A, Stani E, Weyts I. MLDCAT-AP. *MLDCAT-AP* 2025. https://semiceu.github.io/MLDCAT-AP/releases/3.0.0/.

Schnitzer B, Österberg L, Skopa I *et al.* Multi-scale model suggests the trade-off between protein and ATP demand as a driver of metabolic changes during yeast replicative ageing. *PLOS Comput Biol* 2022;**18**(7):e1010261. https://doi.org/10.1371/journal.pcbi.1010261.

Schweidtmann AM, Zhang D, Von Stosch M. A review and perspective on hybrid modeling methodologies. *Digit Chem Eng* 2024;**10**:100136. https://doi.org/10.1016/j.dche.2023.100136.

Sego TJ, König M, Fonseca LL *et al.* EFECT: A Method to Quantify the Reproducibility of Stochastic Simulations, arXiv:2406.16820. Preprint, arXiv, 15 Aug. 2025. https://doi.org/10.48550/arXiv.2406.16820.

Sengar SS, Hasan AB, Kumar S *et al.* Generative Artificial Intelligence: A Systematic Review and Applications, arXiv:2405.11029. Preprint, arXiv, 17 May 2024. https://doi.org/10.48550/arXiv.2405.11029.

Shameer S, Wang Y, Bota P *et al.* A hybrid kinetic and constraint-based model of leaf metabolism allows predictions of metabolic fluxes in different environments. *Plant J* 2022;**109**(1):295–313. https://doi.org/10.1111/tpj.15551.

Sluka JP, Fu X, Swat M *et al.* A Liver-Centric Multiscale Modeling Framework for Xenobiotics. *PLOS ONE* 2016;**11**(9):e0162428. https://doi.org/10.1371/journal.pone.0162428.

Smith LP, Hucka M, Hoops S *et al. SBML Level 3 Package: Hierarchical Model Composition, Version 1 Release 3*. published online 2015. https://doi.org/10.2390/BIECOLL-JIB-2015-268.

Smith LP, Malik-Sheriff RS, Nguyen TVN *et al.* Verification and reproducible curation of the BioModels repository. *PLOS Comput Biol* 2025;**21**(12):e1013239. https://doi.org/10.1371/journal.pcbi.1013239.

Soiland-Reyes S, Sefton P, Crosas M *et al.* Packaging research artefacts with RO-Crate. *Data Sci* 2022;**5**(2):97–138. https://doi.org/10.3233/DS-210053.

the FAIRsharing Community, Sansone SA, McQuilton P *et al.* FAIRsharing as a community approach to standards, repositories and policies. *Nat Biotechnol* 2019;**37**(4):358–67. https://doi.org/10.1038/s41587-019-0080-8.

Thornburg ZR, Maytin A, Kwon J *et al.* Bringing the genetically minimal cell to life on a computer in 4D. *Cell* 2026;**189**(9):2582-2597.e27. https://doi.org/10.1016/j.cell.2026.02.009.

Tiwari K, Kananathan S, Roberts MG *et al.* Reproducibility in systems biology modelling. *Mol Syst Biol* 2021;**17**(2). https://doi.org/10.15252/msb.20209982.

Van Aalst M, Lahlou A, Hassan T *et al.* Web-based collaborative model development in interdisciplinary consortia: Design principles and practical guidance. *PLOS Biol* 2026;**24**(6):e3003825. https://doi.org/10.1371/journal.pbio.3003825.

Van Der Aalst WMP, Bichler M, Heinzl A. Responsible Data Science. *Bus Inf Syst Eng* 2017;**59**(5):311–3. https://doi.org/10.1007/s12599-017-0487-z.

Voit EO. Perspective: Systems biology beyond biology. *Front Syst Biol* 2022;**2**:987135. https://doi.org/10.3389/fsysb.2022.987135.

Von Dassow G, Meir E, Munro EM *et al.* The segment polarity network is a robust developmental module. *Nature* 2000;**406**(6792):188–92. https://doi.org/10.1038/35018085.

Waltemath D, Adams R, Beard DA *et al.* Minimum Information About a Simulation Experiment (MIASE). *PLoS Comput Biol* 2011;**7**(4):e1001122. https://doi.org/10.1371/journal.pcbi.1001122.

Waltemath D, Bergmann F, Adams R *et al.* Simulation Experiment Description Markup Language (SED-ML) : Level 1 Version 1. *Nat Preced* published online 25 Mar. 2011. https://doi.org/10.1038/npre.2011.5846.1.

Waltemath D, Golebiewski M, Blinov ML *et al.* The first 10 years of the international coordination network for standards in systems and synthetic biology (COMBINE). *J Integr Bioinforma* 2020;**17**(2–3):20200005. https://doi.org/10.1515/jib-2020-0005.

Wehling L, Singh G, Mulyadi AW *et al.* Talk2Biomodels: AI agent-based open-source LLM initiative for kinetic biological models. *BMC Bioinformatics* 2025;**26**(1):276. https://doi.org/10.1186/s12859-025-06310-1.

Wilkinson MD, Dumontier M, Aalbersberg IjJ *et al.* The FAIR Guiding Principles for scientific data management and stewardship. *Sci Data* 2016;**3**(1):160018. https://doi.org/10.1038/sdata.2016.18.

Wolkenhauer O, Auffray C, Jaster R *et al.* The road from systems biology to systems medicine. *Pediatr Res* 2013;**73**(2–4):502–7. https://doi.org/10.1038/pr.2013.4.

Wolstencroft K, Krebs O, Snoep JL *et al.* FAIRDOMHub: a repository and collaboration environment for sharing systems biology research. *Nucleic Acids Res* 2017;**45**(D1):D404–7. https://doi.org/10.1093/nar/gkw1032.

Zheng Y, Schupp JC, Adams T *et al.* A deep generative model for deciphering cellular dynamics and in silico drug discovery in complex diseases. *Nat Biomed Eng* 2025;**9**(12):2155–80. https://doi.org/10.1038/s41551-025-01423-7.

Zrimec J, Correa S, Zagorščak M *et al.* Evaluating plant growth–defense trade-offs by modeling the interaction between primary and secondary metabolism. *Proc Natl Acad Sci* 2025;**122**(32):e2502160122. https://doi.org/10.1073/pnas.2502160122.

Zrimec J, Fu X, Muhammad AS *et al.* Controlling gene expression with deep generative design of regulatory DNA. *Nat Commun* 2022;**13**(1):5099. https://doi.org/10.1038/s41467-022-32818-8.